\documentclass[12pt]{article}

\usepackage[english]{babel}

\usepackage{epsfig}
\usepackage{dcolumn}
\usepackage{amsmath}
\usepackage{cite}
\usepackage{changebar}
\usepackage{graphicx}

\usepackage{graphicx}
\usepackage{dcolumn} 
\usepackage{bm}      
\usepackage{umlaut}

\begin{document}

%

\title{\vspace*{-0cm}\hfill {\normalsize\bf BI-TP 2003/21} \\ ~\\
The Effects of Quantum Entropy on the Bag Constant}

\author{David E. Miller$^{1,2}\,$\thanks{dmiller@physik.uni-bielefeld.de,
 $\;\;$ om0@psu.edu}
 $\;$ and Abdel-Nasser Tawfik$^{1,3}\,$\thanks{tawfik@physik.uni-bielefeld.de} \\~\\
 {\small $^1$ Fakult\"at f\"ur Physik, Universit\"at Bielefeld, Postfach
 100131, D-33501
 Bielefeld, Germany} \\ 
 {\small $^2$ Department of Physics, Pennsylvania State University, 
         }
 {\small Hazleton, Pennsylvania 18201, USA}\\
 {\small $^3$ Department of Physics, Faculty of Science, 
              South-Valley University, Sohag, Egypt} }



\date{}
\maketitle

\begin{abstract}
The effects of quantum entropy on the bag constant are studied
at low temperatures and small chemical potentials. The inclusion of the
quantum entropy of the quarks in the equation of state provides the hadronic
bag with an additional heat which causes a decrease in the effective latent
heat inside the bag. We have considered two types of baryonic bags, $\Delta$ and
$\Omega^-$. In both cases we have found that the bag constant without the 
quantum entropy almost does not change with the temperature and the quark
chemical potential. The contribution from the quantum entropy to the
equation of state clearly decreases the value of the bag constant.  
\end{abstract}

\noindent
PACS: 12.39.-x \hspace*{2mm} Phenomenological Quark Models, \\
\hspace*{14mm}05.30.-d \hspace*{1.5mm} Quantum Statistical Mechanics 


\section{\label{sec:1} Introduction}

The entropy is a concept which has taken on many meanings throughout the
sciences. Its usual sense relates the heat changes to the likelihood of the
related processes at various determined temperatures. In the limit of low
temperatures Planck~\cite{Planck64} noted that a mixture of different
substances retained a finite entropy even at absolute zero. This result 
is quite contrary to the usual interpretation of the Nernst heat theorem, 
for which the entropy should vanish in the low temperature limit. 
It was Schr\"odinger~\cite{Schr} who pointed out a similar observation 
for $N$ atoms each with a two level ground state resulting in $2^N$ states. 
In this case there should be an entropy of $N\ln2$ with Boltzmann's
constant $k$ taken to be unity. These results are consistent with the quantum
statistical definition~\cite{vNeu} of the entropy $S$ using the density
matrix $\rho$ which relates directly to the wavefunction. This entropy 
is given by the trace over the quantum states as follows:
\begin{equation}
S = - \hbox{ Tr } \left[ \rho \ln \rho \right] \label{eq:0}.
\end{equation}
Schr\"odinger's result~\cite{Schr} can be readily attained for a system 
of $N$ spin one half states. Although the name "Quantum Entropy" implies
the construction of the density matrix $\rho$ from the quantum states,
the actual mathematical form is well rooted in the laws of classical 
physics~\cite{Planck64}. The entropy of mixing of different types of
ideal gases can be calculated in the same way as Eq. (1) by replacing
$\rho$ with $x_i$, which is just the proportion of each constituent 
type $i$ in the total gas system. Thus the entropy of mixing becomes 
$-{\sum_i}x_i{\ln x_i}$, where the sum has replaced the trace operation.
This expression is clearly a constant independent of the temperature 
so that it must remain at absolute zero~\cite{Planck64}. 

    In recent work we have applied these ideas to the quark singlet ground
state of the hadrons~\cite{Mill}. The color symmetry $SU(3)_c$ provides a 
quantum entropy for each of the colored quarks with the value $\ln3$. The
ground state entropy reflects 
the mixing probabilities inside the hadrons at vanishing temperature. We
have extended this result to models at finite temperatures~\cite{MiTa}. 
In this work we investigate the contribution of this entropy to the 
equation of state for the quarks inside the hadrons using the
phenomenological bag model for strong interactions~\cite{DoGoHo}. We 
assume that inside the hadron bag all of the strong interactions at low
temperature $T$ and small quark chemical potential $\mu_q$ are included in the
quark and gluon condensates. In particular, we will look at a special model
for baryons at low temperatures and small chemical potentials, for which the
effective degrees of freedom are the quark and gluon colors. All other
couplings are taken so that the spin and flavor are not explicitly considered.  

     In the limit of low temperatures the interactions in the ground state 
are expressed in terms of the vacuum expectation values of the quark and 
the gluon fields. The calculation of these vacuum contributions are gotten 
from the operator product expansion using the QCD sum 
rules~\cite{SVZ,ReRuYa}, which has the local operators of
dimension four yielding the main contributions to the
thermodynamics~\cite{Leut,BoMi,Milli}. The pure gluon vacuum expectation
value is calculated~\cite{Nari} from the product of the field strength tensors
\hbox{$G^a_{\mu \nu}G_a^{\mu \nu}$} including the renormalization group beta 
function, where the repeated indices are summed over their range of values.
From hereon we shall simply write it as $<G^2>_0$, which is often called the 
vacuum gluon condensate. This vacuum gluon condensate can be extracted~\cite{SVZ} 
from the charmonium spectrum to yield a consistently estimated~\cite{DoGoHo} 
value of about $1.95\;$GeV/fm$^3$. For the quark condensates
\hbox{$m_q<\bar{q}q>_0$} we consider two extreme cases - the light quarks
\hbox{$m_{lq}<u\bar{u}+d\bar{d}>_0$} and the pure strange quarks,
\hbox{$m_s<\bar{s}s>_0$}, 
vacuum expectation values. We use the accepted value of the pion decay
constant $F_{\pi}$ as $92.4\;$MeV~\cite{DoGoHo} with the respective
average light quark mass $m_{lq}$ of $6\;$MeV and strange quark mass $m_s$ of
$150\;$MeV. With these values together with the values of the average pion
$138\;$MeV and kaon $496\;$MeV masses~\cite{PDG} we find the light
and strange quark condensates respectively to be $-42\;$MeV/fm$^3$ and
$-273\;$MeV/fm$^3$. The averaged vacuum contribution to the fields of 
dimension four in the equation of state can be calculated from the
operator product expansion~\cite{Nari} using \hbox{$<G^2>_0 + m_q<\bar{q}q>_0$}.
Thus we find the estimated vacuum contributions to these two extreme cases 
to be $1.91\;$GeV/fm$^3$ for the light quarks and $1.68\;$GeV/fm$^3$ for the
strange quarks. Furthermore, we remark here that both the gluon and quark 
condensates have the same color singlet ground state $0^{++}$ often 
associated with the scalar glueball state~\cite{HanJoPet}.

   For the present work we have chosen these extreme cases for special 
states of baryonic structure. If we look at the spin $3/2$ structure of 
$\Delta$ and $\Omega^-$ as examples of these ground state structures, 
we get a minimal effect from the spin entanglement and flavor mixing. 
Nevertheless, we include the degeneracy factor due to the quark spins 
for the integration over the momenta. The usual sum over the flavors 
is replaced by a factor of three in the baryons. We keep the degeneracy 
factor due to the gluon spins since both polarizations are possible. 
In the following work we shall use the trace anomaly for the substitution 
of the above extracted values for the vacuum condensates into the equation 
of state. It is known~\cite{Leut,BoMi,Milli} that the temperature has very 
little effect on these values at temperatures well below $100\;$MeV. Thus 
we can look at the quantum effects of the entropy on the bag constant 
${\mathbf B}$ for low temperatures $T$ and small chemical potentials 
$\mu_q$. Here we do not look at the explicit dependence of the bag 
constant on $\mu_q$ at zero temperature.

\section{\label{sec:2} Equation of State}

At finite temperatures and quark chemical potentials we choose the grand canonical
partition function ${\mathcal{Z}}(T,V,\mu_q)$ in order to write down the equation
of state in terms of difference between the energy density and the pressure
\hbox{$\varepsilon(T,\mu_q)-3p(T,\mu_q)$}. This expression may be  
represented in terms of the trace of the energy-momentum tensor averaged 
over these variables \hbox{$\left<\Theta_{\nu}^{\nu}\right>_{T,{\mu_q}}$},
where the repeated indices represent the sum over the Lorentz indices. 
In the formulation of the bag model~\cite{DoGoHo} the thermodynamics is 
usually included with a bag energy density \hbox{$\varepsilon=+\mathbf B$} and a
bag pressure \hbox{$p=-\mathbf B$}, which generally represents the energy density
and the confining pressure of the bag against the vacuum. Hereupon, we may
describe the expectation values for the gluon and quark condensates from
the equation of state as a limiting value which exists in the sense  
\hbox{$\lim_{T\rightarrow
    0}<\Theta_{\nu}^{\nu}>_T\equiv\epsilon-3p$}.  
After applying the first law of thermodynamics relating the internal energy
density to the entropy density, the pressure and the chemical potential we find
the equation of state using the proton volume $V_P$ from the charge
radius~\cite{PDG} with the value $2.76\;$fm$^3$.
\begin{eqnarray}
\left<\Theta_{\nu}^{\nu}\right>_{T,{\mu_q}} &=& T\;\left[\frac{3{\cal
   S}_{q,3}(T)}{V_P} + s(T,\mu_q)\right] - 4\; [p(T,\mu_q) - {\mathbf B}] +
   3\; \mu_q\; \rho(T,\mu_q)\;.
\label{eq:2}
\end{eqnarray}
\noindent
Here we include~\cite{Mill,MiTa} the ground state entropy ${\cal S}_{q,3}$ and 
the bag constant ${\mathbf B}$, which is usually assumed as independent of the
parameters of the ensemble. 
The other thermodynamical quantities in the equation of state are given as follows:

\begin{eqnarray} 
p(T,\mu_q) &=& \frac{3}{\pi^2}\;T\;\int_0^{\infty} k^2\; dk
   \left\{\ln\left[1+e^{-\frac{\epsilon(k)+\mu_q}{T}}\right]+
   \ln\left[1+e^{-\frac{\epsilon(k)-\mu_q}{T}}\right]\right\}+\frac{8\pi^2}{45}T^4,\label{eq:3} \\
\rho(T,\mu_q) &=& \frac{3}{\pi^2}\;\int_0^{\infty} k^2\; dk
  \left\{ \frac{1}{e^{\frac{\epsilon(k)-\mu_q}{T}}+1}  -
  \frac{1}{e^{\frac{\epsilon(k)+\mu_q}{T}}+1}\right\},  \label{eq:4}\\
s(T,\mu_q)&=& \frac{3}{\pi^2}\;\frac{1}{T}\;\int_0^{\infty} k^2\; dk
  \left\{ \frac{\epsilon(k)+\mu_q}{e^{\frac{\epsilon(k)+\mu_q}{T}}+1}  +
  \frac{\epsilon(k)-\mu_q}{e^{\frac{\epsilon(k)-\mu_q}{T}}+1}\right\} + 
  \frac{p_q(T,\mu_q)}{T} + \frac{32\pi^2}{45} T^3,  \label{eq:5}\\
{\cal S}_{q,3}(T) &=& -\frac{1}{3}\left(1-{e^{-m_q/T}}\right)
  \ln\left[\frac{1}{3}\left(1-{e^{-m_q/T}}\right)\right]
  -2 z \left[\ln(z) \cos(\theta)-\theta \sin(\theta)\right],  \label{eq:6}
\end{eqnarray}
\noindent
where $\varepsilon(k)^2=m_q^2+k^2$ is the single particle energy and the
values for $z$ and $\theta$  respectively  are 
\begin{eqnarray}
        z &=& {\frac{1}{3}} \left[1 + e^{-m_q/T}~+~e^{-2 \;
        m_q/T} \right]^{1/2} \;, \nonumber \\
  \theta &=& \arctan{\left(\frac{\sqrt{3} \; e^{-m_q/T}}
          {2 + e^{-m_q/T}}\right)}\;.\label{eq:deftheta}
\end{eqnarray}

\noindent
The equations~\ref{eq:3}, \ref{eq:4}
and \ref{eq:5} give the pressure, the baryon number density and 
the entropy density inside the baryonic bag depending on
$T$ and $\mu_q$ in the usual way. In Eq. \ref{eq:5} the expression
$p_q(T,{\mu}_q)$ means just the quark contribution to the
pressure in Eq. \ref{eq:3}. The effective degrees of freedom 
which we are considering here are merely the quark colors. 
The contributions of gluons to these quantities are also taken into
account, for which we considered the spin degeneracy due to the possible
polarization. However, for the sake of simplicity, we have given the gluon 
radiation in the grand canonical partition function in an
approximated form\footnote[1]{At high temperatures the pressure of the
  equilibrated ideal bosonic gas of gluons reads \begin{eqnarray}
p(T)_{gluon} &=& \frac{8}{45} \pi^2\;T^4\; \left[1- \frac{15 \alpha_s}{4
               \pi} + \cdots\right] \nonumber 
\end{eqnarray} where  $\alpha_s$ is the running strong interaction
coupling constant, which depends on $T$ and $\mu_q$. At low
temperatures and small chemical potentials we can take
\hbox{$\alpha_s\rightarrow 0$}.} 
which, for instance, appears in the second term of Eq.~\ref{eq:3}. This
approximation is appropriate since both $T$ and $\mu_q$ remain small
compared to $\Lambda_{QCD}$.
  
By using the values of vacuum condensates given above, we can numerically 
solve Eq.~\ref{eq:2} to get the dependence of the bag constant 
${\mathbf B}$ upon the temperature $T$ and quark chemical potential $\mu_q$. 
Here we have considered two special baryonic structures of the hadron bag model, 
for which we shall compare the constancy of the bag constant with and without 
the quantum entropy ${\cal S}_{q,3}$. The inclusion of quantum entropy 
term~\cite{Mill,MiTa} given in Eq.~\ref{eq:6} when set into the
equation of state Eq.~\ref{eq:2} has the effect of decreasing the 
pressure inside the bag. This leads to a decrease in the value 
of ${\mathbf B}$ needed to preserve the hadron bag's stability against 
the force of the outside vacuum.

\section{\label{sec:3} Results and Discussion}

\begin{figure}
 {\hspace*{-1cm}\includegraphics[width=10cm]{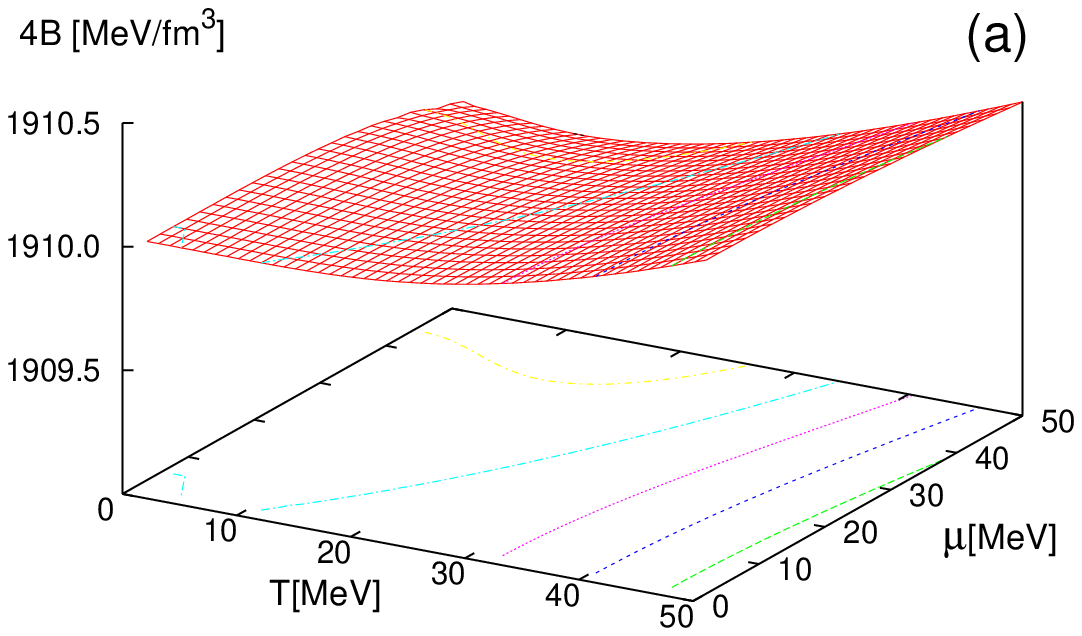}}%
 \vspace*{-1cm}
 {\hspace*{-1cm}\includegraphics[width=10cm]{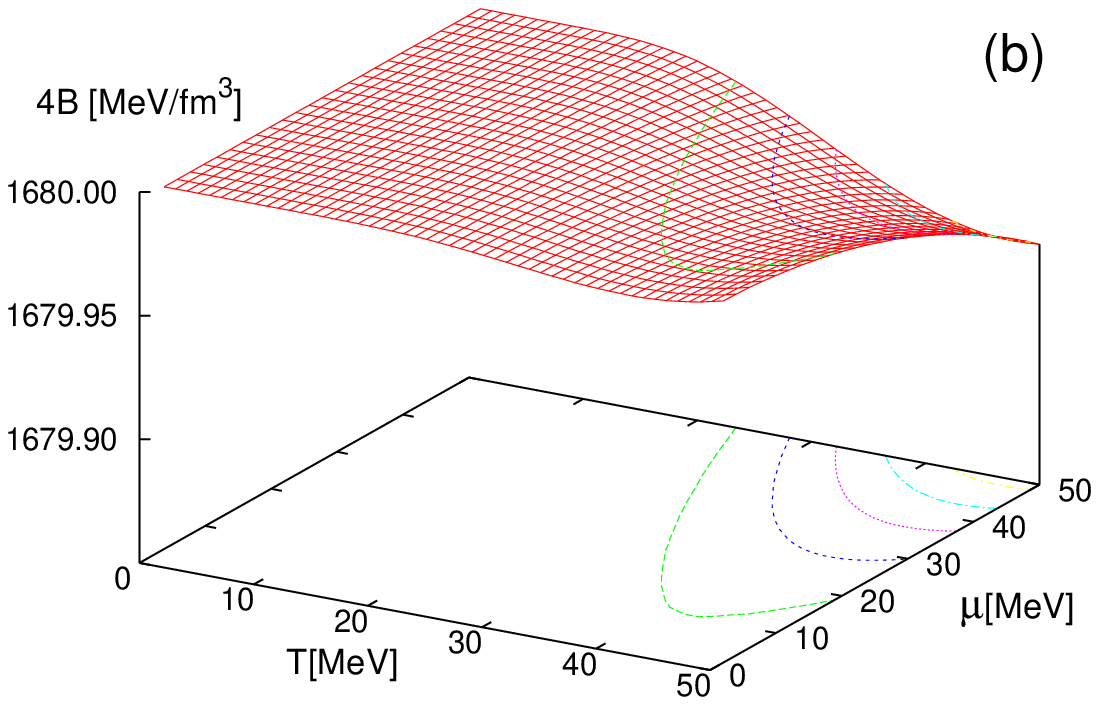}}%
 \caption{\label{fig:1} The  panel (a) depicts $4B$ as a function 
   of $T$ and $\mu_q$ in a system with three light quarks. 
   ${\cal S}(T)_{q,3}$ is not included here in the equation
   of state. The panel (b) gives the same dependence 
   for three strange quarks. }
 \end{figure}
 
 \begin{figure}
 {\hspace*{-1cm}\includegraphics[width=9cm]{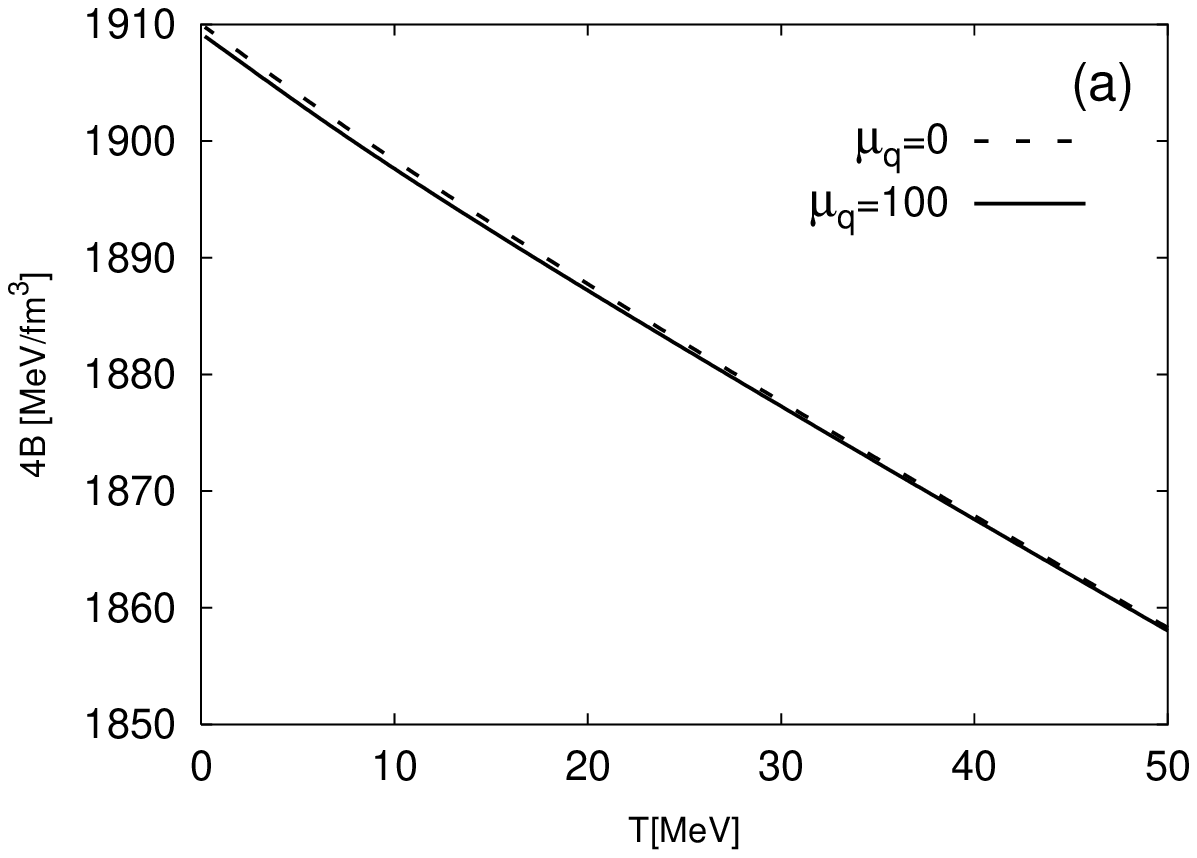}}%
 \hspace*{.5cm}
 {\hspace*{-1cm}\includegraphics[width=9cm]{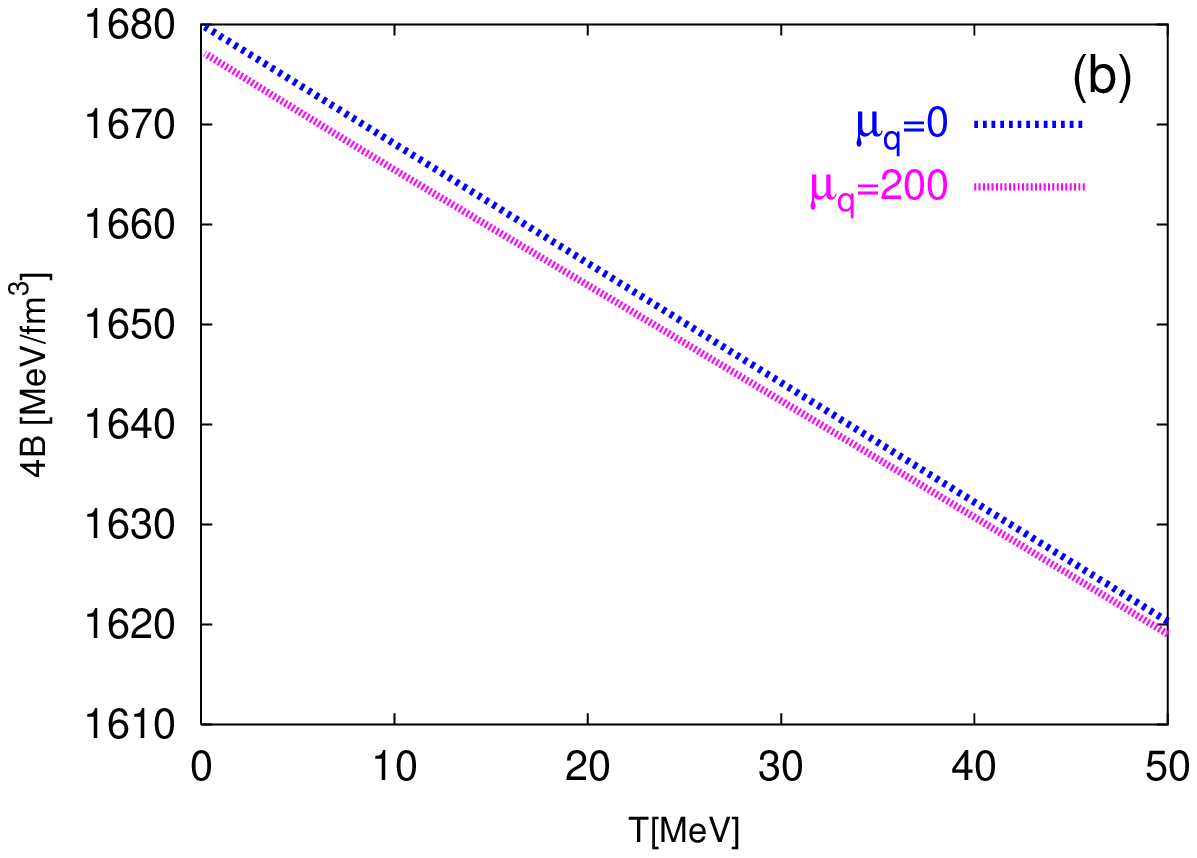}}%
 \caption{\label{fig:2} The panel (a) shows the thermal structure of three
   light quarks in a hadron bag for two values of the chemical potential. 
   The similar thermal structure for three strange quarks in a hadron bag
   appears in (b). Both have the added quantum entropy term
   ${\cal S}(T)_{q,3}$. } 
 \end{figure}

In Fig.~\ref{fig:1}(a) we plot $4{\mathbf B}$ as a function of both $T$
and $\mu_q$ for the baryonic state $\Delta$ given in Eq.~\ref{eq:2} while
leaving out the quantum entropy contribution~\cite{Mill,MiTa}. Here we
have used the vacuum expectation value $1.91\;$GeV/fm$^3$ for the light quarks. 
We notice - at least for the region of $T$ and $\mu_q$ 
which we are considering in Fig.~\ref{fig:1} up to $50\;$MeV - that 
$4{\mathbf B}$ almost stays constant at the very low temperatures. 
Afterwards it only slightly increases with $T$. However, the
dependence on $\mu_q$ is in comparison quite weak. For
relatively large $\mu_q$ and very small temperature $4{\mathbf B}$ remains
constant. For increasing values of $\mu_q$ the size of $4{\mathbf B}$
decreases. For larger $T$ the decrease due to $\mu_q$ will be moderated.

     In Fig.~\ref{fig:1}(b) we show the corresponding results for he baryonic
state $\Omega$ without the quantum entropy. Here we have used the vacuum 
expectation value $1.68\;$GeV/fm$^3$ for the strange quarks. In this case
we note the range of both $T$ and $\mu_q$ in which $4{\mathbf B}$ 
remains constant is much wider than in the case of $\Delta$. 
This fact is due to the strange quark mass, which is very much heavier 
than the masses of light quarks. Beyond this range it decreases only 
marginally. Thereafter it starts to increase in a region not shown here. 
For larger $T$ and $\mu_q$ we could expect that $4{\mathbf B}$ in the 
$\Omega$ bag should increase in a way similar to the $\Delta$ bag. 
Thus we may conclude that $4{\mathbf B}$ is almost constant, especially
at very low temperatures and very small quark chemical potentials, 
when we do not consider the quantum entropy in the equation of state. 
Similarly we note that the dependence on $\mu_q$ is small but not fully
negligible even though at higher values of $T$ it seems not depending 
on $\mu_q$. In the high temperature region we expect that the values of
\hbox{$\epsilon(T)-3p(T)$} do not vanish for which the gluonic radiation 
contributions become very dominant. Thus at vanishing temperatures
\hbox{$\epsilon-3p$} approaches $4B$.
We may consider the region of small $T$ and $\mu_q$ in which the bag
constant remains relatively unchanged, which for the strange quarks is
much wider.  

     The inclusion of the quantum entropy density leads to quite different 
qualitative effects which may be seen in Fig.~\ref{fig:2}. 
Here we note an almost linear decrease of $4{\mathbf B}$ with increasing $T$. 
The dependence on both $T$ and $\mu_q$ is in $\Omega$ bags stronger than that 
in $\Delta$ bags. We should emphasize that the changes in $4{\mathbf B}$ 
from $\mu_q$ is a gradually varying one with a  decrease in the slope. 
This is quite different from $T$ dependence, for which 
$4{\mathbf B}$ is almost linearly decreasing function. 
One reason for the rather weak dependence on $\mu_q$ is that the 
quantum entropy term, ${\cal S}$,  which is not included in the 
grand canonical partition function, has only been given a dependence on $T$. 
We may expect that this behavior will be valid for a much wider 
range of temperature, perhaps even up to $T_c$.

\section{\label{sec:4} Conclusion and Outlook}

Finally we are able to determine that the presence of the quantum entropy 
arising from the $SU(3)_c$ color symmetry in the equation of state
provides a strong temperature dependence for the bag constant ${\mathbf B}$.
However, the equation of state yields a much weaker dependence of
${\mathbf B}$ on the quark chemical potential $\mu_q$, from which
${\mathbf B}$ is changed very slowly at low temperatures. The
inclusion of the quantum entropy term in the equation of state
for the two considered baryonic structures leads to an acceleration
of the decrease in ${\mathbf B}$ with increasing temperatures has
an approximately linear decline. We have contrasted these results in
both these cases to the same properties without the quantum entropy.

     Based on these results we shall plan further studies into the behavior 
of the structure of confined quark matter at low temperatures and 
small quark chemical potentials. A further consideration of the idea that
glueballs could appear as $0^{++}$ state in a Bose-Einstein 
condensate~\cite{HanJoPet} seems to be a quite promising further point.
The importance of the glueball degrees of freedom for describing the 
hadronic phase for temperature below $T_c$ has already been investigated
~\cite{KRT1,KRT2}. Also the existence of spin-color waves in the bag
we would like to study further.

~\\
{\bf Acknowledgments}\\
We wish to acknowledge the stimulating discussions with Frithjof~Karsch,  
Krzysztof~Redlich and Helmut~Satz. D.E.M. is very grateful to the
Pennsylvania State University Hazleton for the sabbatical leave of absence
and to the Fakult\"at f\"ur Physik der Universit\"at Bielefeld. 


\end{document}